# Growth mechanisms of GaN/GaAs nanostructures by droplet epitaxy explained by complementary experiments and simulations.

*Guy Tsamo[1], Alla G. Nastovjak[2, 3], Nataliya L. Shwartz[2, 3], Philip E. Hoggan[1], Christine Robert-Goumet[1], Alberto Pimpinelli[1], Matthieu Petit[4], Alain Ranguis[4], Emmanuel Gardes[5], Mamour Sall[6], Luc Bideux[1]and Guillaume Monier[1]*

[1]Université Clermont Auvergne, Clermont Auvergne INP, CNRS, Institut Pascal UMR 6602, F-63000 Clermont-Ferrand, France

[2]A.V. Rzhanov Institute of Semiconductor Physics SB RAS, 13 Lavrentiev aven., Novosibirsk, 630090, Russia

[3] Novosibirsk State Technical University, 20 K. Marx str., Novosibirsk, 630073, Russia

[4]Aix Marseille Université, CNRS, CINaM UMR 7325, 13288, Marseille, France

[5]Université Clermont Auvergne, CNRS, IRD, OPGC, Laboratoire Magmas et Volcans, F-63000 Clermont-Ferrand, France

[6]CIMAP Normandie Univ, CEA, CNRS, ENSICAEN, UNICAEN, 14000 Caen, France

ABSTRACT

In this work, we present conception and study of gallium nitride (GaN) nanostructures on a gallium arsenide (GaAs) substrate with (111)A orientation. The nanostructures were designed by GaN droplet epitaxy and studied *in-situ* by X-ray photoelectron spectroscopy and *ex-situ* by atomic force microscopy, scanning electron microscopy and transmission electron microscopy. These studies were coupled with kinetic Monte Carlo simulations to precisely understand the phenomena occurring during the nitridation and to find the optimum conditions for complete nitridation of gallium droplets. The HRTEM observation showed a cubic (zinc blende) crystal structure of the GaN nanodots for a nitridation at 300°C. Ramping the temperature from 100°C to 350°C during droplet nitridation enabled to obtain a very high density (>$10^{11}$cm$^{-2}$) of GaN nanodots with the zinc blende crystallinity.

# 1  Introduction

Molecular beam epitaxy (MBE) is an established method of quantum dot (QD) growth. Quantum dot formation in lattice-mismatched material systems mainly uses Stranski-Krastanov growth. In this case, quantum dots grow on a 2D wetting layer. The lattice mismatch between substrate and wetting layers, leads to island formation that relax the elastic stress, albeit incompletely, giving dislocations or structural defects in QDs, due to vertical deformations [1].

An alternative approach is the so-called droplet epitaxy technique [2,3,4]. It was first reported in 1991 by Koguchi *et al*. The first step is to deposit the group III metal element, for example Ga, In or Ga+In, to form droplets on the substrate. The second step is to supply group V element to transform the metal droplets into III-V QDs. These quantum dots are used to build devices such as photovoltaic solar cells, light emitting diodes, single photon generators, entangled photon generators, bipolar field effect transistors, biomedical sensors, photodetectors, photodiodes [5,6].

The work presented here uses gallium arsenide substrate, with a surface orientation of (111)A to grow GaN nanostructures by droplet epitaxy. GaAs has the advantage of a higher direct gap and charge carrier mobility than silicon. It is therefore of interest for optoelectronic applications. As the temperature at which GaAs evaporation begins is around 550°C for the (111)A orientation, it is imperative to crystallize below this temperature to avoid melting it and to avoid Ga atoms to diffuse. Hence, droplet epitaxy on GaAs is done at relatively low temperature [7–10]. In the specific case of GaN/GaAs(111)A dots, these temperatures between 100°C-350°C result in little nitridation of the GaAs substrate [11]. Dissolution of the element V in the metal may perturb and cause instability in the droplet, resulting in the formation of a III-V nanostructure with undissolved metallic gallium in the III-V matrix of the nanostructure. Here a core-shell nanostructure is formed. Formation of such nanostructures has already been observed [12–14].

GaN is a semiconductor with a large direct gap (3.2-3.4 eV) that enables light to be emitted and absorbed in the near-ultraviolet range. The design of nanostructures of this semiconductor extends its field of application to quantum computing [15–22]. For optimum use of the nanostructures produced, the main objective is to obtain crystalline nanostructures that are homogeneous in size and distribution while maintaining a high density.

To optimize the growth parameters of GaN nanostructures fabricated by droplet epitaxy on GaAs(111)A substrates, we combine X-ray photoelectron spectroscopy (XPS), scanning electron microscopy (SEM), atomic force microscopy (AFM) and high-resolution transmission electron microscopy (HRTEM) experiments with Kinetic Monte Carlo (KMC) and Density Functional Theory (DFT) simulations.

## 2   Experimental Methods

Nanostructure growth takes place in an ultra-high vacuum (UHV) frame made up of three chambers: the introduction chamber connected to the preparation chamber and the analysis chamber, all separated by valves. Samples are placed in one or other of the chambers using a transfer rod, keeping samples in UHV. A Knudsen-type metallic gallium deposition cell connected to the preparation chamber was calibrated using a quartz crystal microbalance. The deposition rate of 0.1 ML/s was achieved for a cell heated at 900 °C (ML=monolayer).

Before gallium deposition, the GaAs(111)A substrate is cleaned in a chemical solution of hydrochloric acid and isopropanol and quickly introduced into the introduction chamber. The substrate is then heated to 530 °C under UHV. This treatment produces a clean surface free of impurities (no oxygen nor carbon peak was detected by XPS). After the cleaning procedure, the recorded XPS peaks As3d and Ga3d each present only one component ascribed to Ga-As bonds at 41.1 eV and 19.3 eV, respectively. The XPS modelling of GaAs with Ga-rich termination has already been studied in reference [23]. The surface XPS ratio Ga3d/As3d of 0.77 shows a 1.5 ML-thick Ga-rich surface, consistent with the (2×2) Ga-rich surface reconstruction [24,25].

A commercial electron resonance cyclotron plasma source (SPECS MPS-ECR), operating in atom beam mode at a pressure of $7 \times 10^{-5}$ mbar (sample current density measured around 10 nA/cm2) was used for the nitridation process.

X-ray photoelectron spectroscopy (XPS) measurements were performed *in situ* after all steps of droplet epitaxy in an ultrahigh vacuum photoelectron spectrometer equipped with an Omicron DAR 400 X-ray source and an Omicron EA 125 hemispherical analyzer calibrated using the method described in reference [26]. The Mg Kα source (1253.6 eV) running at 300 W is located at an angle of 55° to the analyzer. A high magnification mode, constant pass energy of the analyzer equal to 20 eV and an energy step of 0.1 eV were used for analysis. Photoelectrons were recorded at normal detection. The authors have developed modelling of the XPS Ga3d, As3d and N1s transitions signals during the different phases: cleaning, metal deposition and nitridating in the reference [27]. These models are based on a geometrical description of the drops by spherical caps (diameter d and height h).

Ga droplets and GaN clusters on GaAs (111)A were imaged *ex situ* with a scanning electron microscope (SEM) and an atomic force microscope (AFM). The SEM is carried out with a Zeiss Supra 55 VP in secondary emission with the Inlens detector. The AFM is a Nanoscope IIIA Multimode (Digital instruments) equipped with a 10 μm × 10 μm × 2.5 mm scanner running in the tapping mode at room temperature with a silicon nitride probe (HQ: NSC15/ALBS, Mikromasch). The curvature radius of the silicon tips was about 10 nm (from the supplier specifications). SEM and AFM images were processed using ImageJ and Gwyddion software, respectively.

A SEM equipped with a focused Xe-ion beam gun (FIB) ThermoScientific Helios 5 PFIB CXe was used to extract electron-transparent thin foils perpendicular to the surface of the samples. Those were characterized by high-resolution transmission electron microscopy (HRTEM) using a JEOL-F200 microscope with a cold field emission electron gun operating at 200 kV on TEM lamellas prepared using the FIB technique.

# 3 Theoretical method

Kinetic Monte Carlo (KMC) simulations were carried out for the growth of GaN nanostructures by droplet epitaxy. Simulation used the SilSim3D [28] program package. This software provides Monte Carlo techniques and was designed to simulate growth and annealing processes during nanostructure formation. KMC simulation of GaAs/GaAs(111)A nanostructures growth by droplet epitaxy has already been studied in the reference [29]. The simulated space is a three-dimensional set of lattice sites arranged with the zinc blende (zb) crystal structure. Each site can be occupied by a particle (atom or molecule) or can be empty.

The model substrate is semi-infinite in the z-direction and cyclic boundary conditions are applied in x- and y-directions. The space above the atom-filled layer consists of empty lattice sites. The following elementary events can occur: particle adsorption, diffusion hop, desorption, chemical reaction. The probability of each event $P$ (except for adsorption) is given by its activation energy $E_a$, which is the input model parameter. Adsorption is a non-activated process and is determined by deposition rate. Particle deposition flux on the substrate surface occurs at fixed angle (in MBE) or at arbitrary one (in CVD) at any surface site. All these events result in rearrangement of particles and change in surface morphology. The probability of a diffusion hop is determined by its activation energy $E_{dif}$ and a pre-exponential factor. The pre-exponential factor of the diffusion hop is equal to the Debye frequency $v = 10^{13}$ s$^{-1}$. The value of the activation energy $E_{dif}(q)$ for atom q is mainly determined by the total binding energy of this atom with its neighbor in the first coordination sphere at the initial site. This is corrected by taking into account the atom neighborhood for initial and final sites of the diffusing atom. Thus, the influence of neighbors in the second coordination sphere is included. A diffusion hop is possible to any unoccupied lattice site within three coordination spheres. The diffusion hop of an atom occurs in two stages: atom exit from its site and its incorporation into a neighboring unoccupied lattice site. Each of the stages is characterized by an independent set of energy parameters. If a particle diffuses to a region where it is not bound to other particles, then this particle can either return to the surface to an arbitrary position within the three coordination spheres or desorb (and be removed from the system). The restriction to three coordination spheres is due to the specific features of the diamond-like cubic lattice structure. When moving from one surface site to a neighboring site, the particle hops into the second coordination sphere relative to its initial position. To cross the step edge, the particle should be able to achieve a diffusive hop into the third coordination sphere. The particle returns to the crystal surface unless it overcomes an additional barrier to particle removal from the system $E_{des\_vac}(q)$. This additional barrier mimics the collective effect of all substrate particles on the evaporating particle. The sum of the barrier for a diffusion hop and that for particle removal from the system constitute the desorption activation energy: $E_{des}(q) = E_{dif}(q) + E_{des\_vac}(q)$, which is the main desorption characteristic of particle q.

The general form of chemical reactions which can occur in the model is A + B → C+ D, where A, B are reagents and C, D the reaction products. The reaction activation energy depends on reagent surroundings. The energy barrier ($E_r$) for a chemical transformation has the form: $E_r = E_{react} + \Delta E_A + \Delta E_B$, where $E_{react}$ is the activation energy of the chemical transformation, and $\Delta E_A$ and $\Delta E_B$ are energy corrections depending on the environment of reactants A and B. The values of the corrections $\Delta E_A$ and $\Delta E_B$ depend on the type and neighborhood of reacting particles. Reactions in the model can take place on the surface or in the volume of the model crystal. Reactions in the gas phase are not considered. Inside the droplet, all lattice sites are occupied, and ordinary diffusion is impossible. Using the reaction: A + B → B + A one can simulate bulk diffusion process when A and B atoms exchange positions. Nitrogen diffuses inside the Ga droplet using this type of reaction. The reaction of the type A + B → C allows us to simulate the formation of a molecule from two atoms.

To simulate GaN nanocluster formation on GaAs(111)A substrate the 5-component system was considered (Fig.1): As, Ga(s), Ga(liq), N, $N_2$ – arsenic in atomic form, gallium in solid and liquid states, nitrogen in atomic and molecular forms, respectively. The following events were taken into account in the model: adsorption of Ga(s) and N atoms, desorption of Ga(s) and Ga(liq) atoms and $N_2$ molecules, diffusion of all types of particle, arsenic, nitrogen and gallium dissolution and diffusion in liquid gallium, GaAs and GaN crystallization. Nitrogen adsorption was considered in atomic form as in plasma assisted MBE. Nitrogen desorption was in $N_2$ molecular form. $N_2$ molecules resulted from the recombination reaction N + N → $N_2$, which is assumed barrier free and irreversible. $N_2$ desorption was assumed to be barrier-free to provide extremely high volatility of diatomic nitrogen. The description of GaAs-Ga(liq) system realization in lattice MC model and energy parameters for this system can be found in [30].

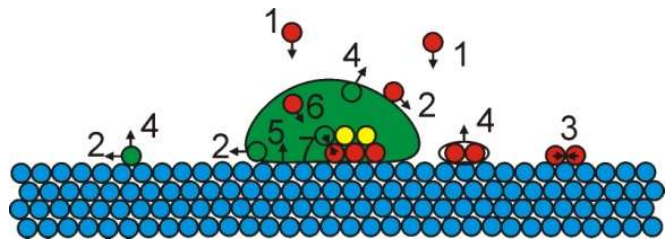

**Figure 1.** The scheme of the model system and considered elementary events: 1 – particle adsorption; 2 – diffusion; 3 – nitrogen recombination N + N → $N_2$; 4 – evaporation; 5 – Ga dissolution into liquid Ga; 6 – nitrogen dissolution and diffusion through the Ga droplet; 7 – GaN crystallization. GaAs substrate is marked by blue, Ga (solid) is marked by yellow and Ga (liquid) by green. Incoming nitrogen atoms are shown in red.

The model makes the following assumptions: all particles of the system are arranged in a zb lattice, including liquid gallium and GaN material. Liquid phase modeling is based on a crystal structure approximation is described in detail in [29]. Elastic deformations are not taken into account.

The key energy parameters for GaN-GaAs(111)A system are binding energies between Ga-Ga, Ga-As, N-N and Ga-N atoms. The As-N interaction is of no great importance due to Ga polarity of the substrate considered. Binding energies between components of the system were set using binding energy values for bulk materials. The energies for homopolar bonds were taken from [31]. Because each particle in this model system has four neighbors, binding energy between two i atoms $E_{i-i}$ is equal to ¼ of bulk binding energy. The binding energies for heteropolar bonds A-B were calculated using an empirical expression [32]:

$$E_{A-B} = \frac{1}{2}(E_{A-A} + E_{B-B}) + 100(\chi_A - \chi_B)^2,$$ where $E_{A-A}$ and $E_{B-B}$ are bulk binding energies of A and B material, $\chi_A$, $\chi_B$ are electronegativities of A and B atoms. Hence, the following energies were used to simulate the full GaN-GaAs(111)A system: $E_{Ga-Ga} = 0.7$ eV, $E_{N-N} = 1.2$ eV, $E_{As-As} = 0.7$ eV, $E_{Ga-N} = 3.2$ eV. Note that energy parameters for liquid and solid gallium were set identical, apart from surface tension. The binding energies between solid gallium atoms obey the quasi-chemical approximation when atom binding energy with the neighbors is proportional to their amount. As to liquid gallium, to correctly mimic liquid drop shape, the binding energies of Ga atom with two or three neighbors were set almost equal: $E_{Ga-2Ga} \approx E_{Ga-3Ga}$, where $E_{Ga-2Ga}$ –is the binding energy of Ga atom with two Ga atoms, $E_{Ga-3Ga}$ – with three Ga atoms. Such a method allows our model to approach the real system: Ga droplets have spherical shape whereas a crystal is faceted.

To calculate lateral and vertical sizes of clusters we measure these parameters in lattice constants and then multiply them by the lattice constant of GaN in its zinc blende phase.

To evaluate the activation energies of GaN dissolution and crystallization, we simulate the solubility of nitrogen in the liquid Ga at different temperatures by KMC. The initial system was a GaN substrate covered with a liquid gallium layer. To prevent the substance from evaporating on the top of the liquid layer, a thin layer of an inert material acts as a cap. During annealing the substrate dissolves in liquid gallium. Annealing continued until nitrogen concentration in the Ga became constant, ~~and~~ i.e .equilibrium was reached. This additional computational experiment allowed us to determine the activation energies of GaN substrate dissolution. Crystallization of GaN in the model is carried out via the reaction Ga(liq) + N → Ga(s) + N. The average activation energy of this process is 0.95 eV, taking the reagent neighborhood into account. The reaction: N + Ga(liq) → Ga(liq) + N simulates the nitrogen dissolution in gallium drop with activation energy 1.3-1.6 eV depending on reagent neighborhood. Using these values, we evaluated equilibrium concentration of nitrogen at low temperatures to be used for following

Ga droplet nitridation. The nitrogen dissolution in the gallium is almost absent at annealing temperatures T=100 °C. In temperature range 200-300 °C equilibrium nitrogen concentration in liquid Ga does not exceed 1%. A concentration of 0,01851 at 227 °C was reported in [33]. This value is well within the concentration values obtained in simulation. The nitrogen concentration in liquid gallium increases with temperature.

## 4   Results and discussion

### 4.1   Gallium deposition on GaAs(111)A substrates

The first step in the design of GaN nanostructures by droplet epitaxy is the deposition of metallic gallium. There is neither a buffer layer grown on the substrate nor GaAs(111) nitridation before droplet deposition. The deposition parameters of gallium at a rate of 0.1 ML/s for 30 s (equivalent to 3 ML) on a GaAs(111)A substrate at a temperature of 300 °C were chosen in accordance with previous studies [10,34,35], to obtain a high droplet density of >10[11] cm-2.

The XPS peak decompositions are shown in the Fig.2. The Ga3d peak shows GaAs (19.3 eV) and Ga-Ga (18.6 eV) bonds are present. This confirms the presence of metallic gallium on the GaAs (111) A substrate. The As3d peak is composed of only Ga-As contribution (41.1 eV).

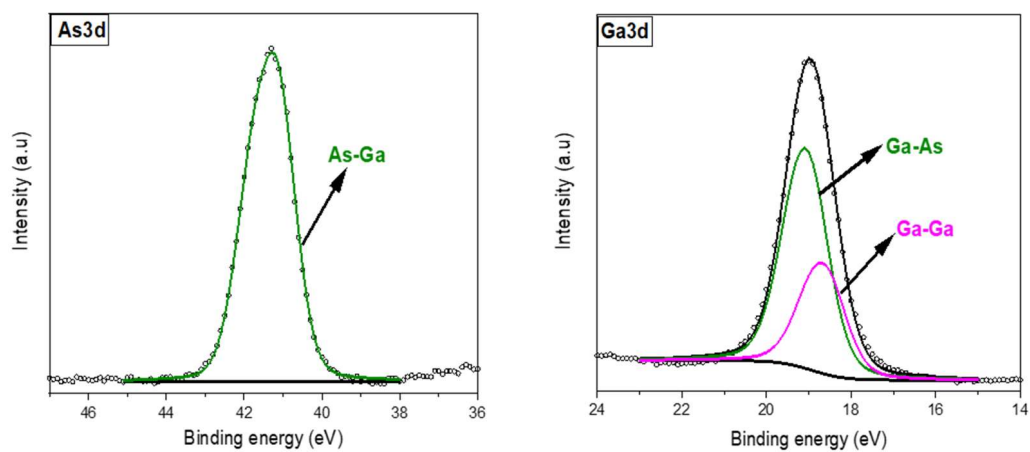

**Figure 2.** XPS peak decompositions after depositing 3 ML of Ga on GaAs (111)A.

Our previous work on XPS signal modelling [27] allowed morphology (coverage, diameter and height) determination of the gallium droplets by comparing calculated and measured ratios. Thus, the ratio of the total intensities of Ga3d and As3d peaks ($\frac{I_{Ga3d}}{I_{As3d}} = 0.95$), the ratio of the Ga-As component of Ga3d peak by the intensity of As3d peak ($\frac{I_{Ga-As}}{I_{As3d}} = 0.77$), and the ratio of the Ga-Ga component of Ga3d peak by the Ga-As component of the Ga3d peak ($\frac{I_{Ga-Ga}}{I_{Ga-As}} = 0.45$) have been calculated after the deposition of 3ML of gallium metal. Moreover, *ex-situ* characterizations have been performed by Scanning Electron (SEM) and Atomic Force (AFM) microscopies (Fig.3). The droplet density, the mean diameter and the mean height are found to be $1.3 \times 10^{11}$ cm$^{-2}$, $17 \pm 3$ nm and $6 \pm 1$ nm, respectively. All the morphological results are presented in the Table 1 and are consistent whatever the used characterization technique.

All the morphological parameters of Ga droplets remained unchanged after a 15 min. annealing at 300°C. In particular, no Ostwald ripening of Ga droplets is observed at this temperature.

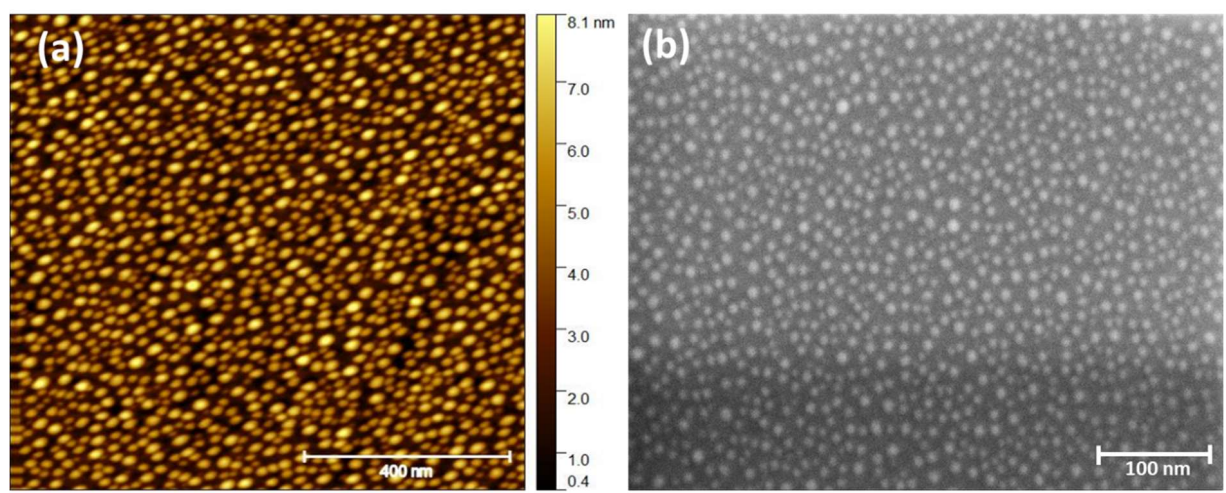

**Figure 3.** AFM (a) and SEM (b) images taken on Ga/GaAs(111)A after 3ML Ga deposition at T=300 °C.

Based on microscopy images, diameter and height distributions of metallic gallium droplets have been calculated. Results are presented in the Fig.4 and Table 1. The diameter histogram indicates a distribution centred on a diameter of around 17 nm. This value is in perfect agreement with the results of the XPS modelling. The same agreement is observed for the drop heights (around 5 nm). Low standard deviations ($\sigma$) in diameter and height show that the nanostructure sizes remain tightly around a mean value. This accounts for the uniformity of the gallium droplets.

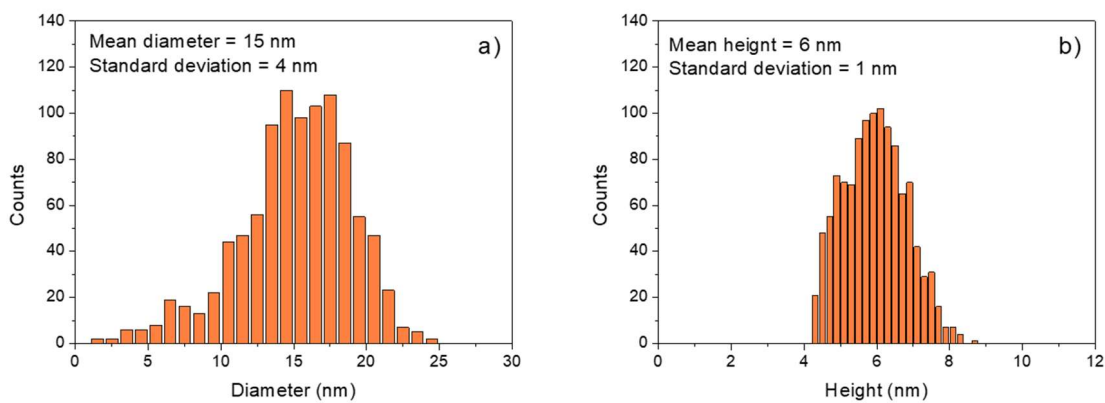

**Figure 4.** Droplet diameter (a) (from SEM) and height (b) (from AFM) distributions after 3 ML Ga depositing.

**Table 1.** Morphological characteristics of Ga/GaAs(111A).

| Morphological characteristics | Density ($\times 10^{10}$ cm$^{-2}$) | Diameter d (nm) | Height h (nm) | Coverage | Contact angle $\beta$ (°) |
|---|---|---|---|---|---|
| Calculated values with XPS | | 16 | 4-5 | 0.4-0.5 | 55-60 |
| Data from AFM and SEM | $13 \pm 2$ | AFM : $17 \pm 3$ SEM : $15 \pm 2$ | AFM : $6 \pm 1$ | 0.4-0.5 | |

To estimate activation energy for gallium diffusion on the GaAs(111)A surface, a KMC simulation of Ga deposition on GaAs substrate was performed. The model substrate size was $200 \times 200$ atomic sites. As for the experiment, a 3 ML of gallium was deposited at the rate of 0.1 ML/s for 30 s. The substrate temperature was set at 300 °C. The simulation was carried out for several values of gallium diffusion barrier energy $E_{diff}$ on GaAs(111)A. Simulation results are shown in Fig.5. Gallium deposition on GaAs(111)A results in gallium droplet formation.

Droplet density and size depend on $E_{\text{diff}}$ value. It should be noted that as $E_{\text{diff}}$ decreases, the shape of the droplets becomes rugged (Fig. 5a). This effect occurs at low temperature and is kinetic effect. The decrease of $E_{\text{diff}}$ leads to an increase in the gallium adatom diffusion length on the surface, and, consequently, to an increase in accumulation by diffusion of gallium in each drop, promoting the development of instability shown by forming distorted drop shapes.

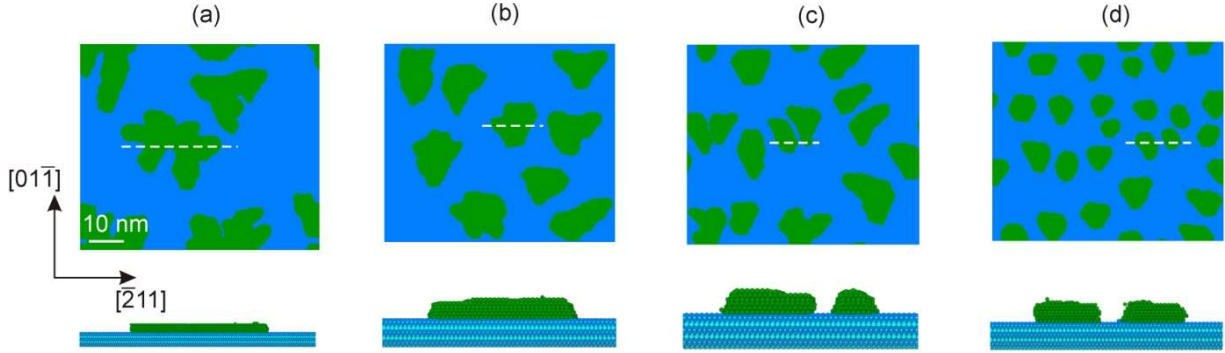

**Figure 5.** Monte Carlo simulation results: Model substrate top views and cross-sections along dashed line after 3 ML Ga deposition during 30 s on GaAs(111)A at T=300 °C as a function of Ga diffusion barrier energies $E_{\text{diff}}$: (a) 0.6 eV; (b) 0.7 eV; (c) 0.8 eV; (d) 0.9 eV. Substrate Ga(solid) and As(solid) atoms are marked in blue and cyan, respectively. Ga(liquid) is marked in green.

The morphological characteristics of Monte Carlo simulation after 3 ML Ga deposition on GaAs(111)A are presented in the table 2. The droplet density in the table is averaged over 5 calculations. The droplet diameter is defined as mean droplet size in the x- and y- directions, which correspond to $[01\bar{1}]$ and $[\bar{2}11]$ directions.

**Table 2.** Morphological characteristics of model gallium droplet after 3 ML Ga deposition on GaAs(111)A.

| Diffusion energy barrier $E_{\text{diff}}$ (eV) | 0.6 | 0.7 | 0.8 | 0.9 |
|---|---|---|---|---|
| Droplet density (cm$^{-2}$) | $6.1 \cdot 10^{10}$ | $1.4 \cdot 10^{11}$ | $2.4 \cdot 10^{11}$ | $4.6 \cdot 10^{11}$ |
| Diameter (nm) | 25.8 | 16.6 | 12.3 | 9.0 |
| Height (nm) | 2.0 | 2.0 | 2.0 | 2.2 |

The gallium diffusion barrier energy across GaAs(111)A surface of 0.7 eV provides an average droplet diameter of 16.6 nm that best approximates the experimental system. For this energy, the density and size of the gallium droplets agree with the experimental values calculated from the SEM and AFM images. It should be noted that the model density of islands was estimated on the unreconstructed GaAs(111)A surface, since the surface reconstructions are not explicitly considered in the model. However, since the droplet density was compared with the real reconstructed surface, the diffusion activation energy determined corresponds to the diffusion on the real reconstructed surface. We thus implicitly take reconstruction of the real surface into account.

This diffusion barrier energy was also determined using DFT calculations as specified in [11]. It was theoretically expected and experimentally confirmed that the GaAs truncated surface (111) A contains a Ga vacancy for every four Ga atoms [36]. This Ga vacancy shows three As atoms forming an As tripod bonded to the upper layer Ga atoms. Periodic boundary conditions were applied to a supercell consisting of 2x2 unit cells in the horizontal plane with a Ga-vacancy (at the center of the 111A plane exposed). Nudged elastic band (NEB) DFT ab initio dynamics calculations [37] of Ga-atom trajectories across this surface (conducted with the VASP software [38,39]), show a potential energy profile with a barrier height of 0.7±0.05 eV, which is consistent with the KMC value of 0.7 eV. The energy along an optimal trajectory as a function of the relative position of the propagating gallium atom is shown in Fig.6. The diffusion trajectory plotted in this figure uses an option of VASP software for selecting initial velocity vector magnitude and direction, for each atom in the supercell. The fact that diffusing Ga on the Ga-rich face is initially stabilised, as the 'probe' Ga atom moves away from the initial position above a surface Ga-atom, towards the vacancy where it is stabilised, is realistic. It proceeds after a low barrier due to repulsion from neighbouring Ga-atoms in the surface. Clearly, if the Ga were instead to begin from the stable vacancy position it would require about 2 eV activation to reach the initial position. The insert shows the top view of a reconstructed (2x2) Ga-rich GaAs(111)A surface constructed with a Ga vacancy as described in the reference [36]. The propagating gallium atom is shown in green.

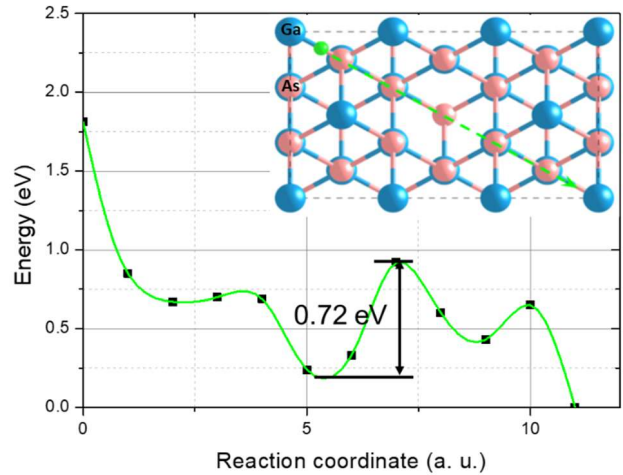

**Figure 6.** Energy as a function of the relative position of the diffusing gallium atom.

## 4.2 Gallium droplet nitridation

The metallic gallium droplets are then exposed to nitrogen flux at $7 \times 10^{-5}$ mbar pressure. Different conditions of nitridation were used: (i) one-step nitridation regime was performed: GaN150 was nitridated at a temperature of 150°C for 30 min and GaN300 was nitridated at a temperature of 300 °C for 15 min and (ii) a stepped nitridation regime was performed: the sample named GaN100→350 was nitridated by ramping up the temperature to 100°C for 15 min, 200 °C for 15 min, 300 °C for 15 min and 350 °C for 15 min. After nitridation, GaN300 and GaN100→350 were annealed at 500 °C for 15 min.

*In-situ* XPS measurements were performed for each sample after nitridation. XPS peak decomposition of the sample GaN150 is presented in Fig.7 as an example. The As3d peak shows As-Ga bonds with a binding energy of 41.1 eV, a free arsenic ($As^0$) component shifted by + 1.5 eV from As-Ga bond, due to the low nitridation of the substrate [11], and an As-O contribution shifted by +2.7 eV with respect to the As-Ga bond. The Ga3d peak decomposition shows three contributions: the Ga-As at 19.3 eV, the Ga-Ga at 18.6 eV and the Ga-N at 20.2 eV. A weak N2s peak is observed at 17.2 eV. The N1s peak shows the N-Ga contribution at 397.2 eV and a low N-O contribution shifted by + 1.5 eV with respect to the N-Ga bonds. The detection of a molybdenum component shifted by -3 eV with respect to the N-Ga bond comes from the sample holder made of molybdenum. The presence of As-O and N-O bonds is due to minor oxygen contamination during nitridation.

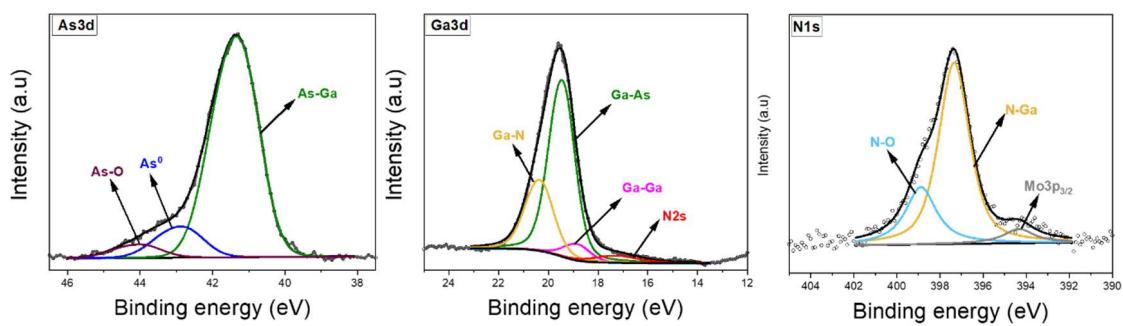

**Figure 7.** XPS peak decompositions after gallium droplet nitridation at 150 °C for 30 min (sample GaN150).

These decomposition parameters can be applied to all the nitridated samples and allow the calculation of the three XPS ratios $\frac{I_{Ga3d}}{I_{As3d}}$, $\frac{I_{Ga-N(Ga3d)}}{I_{As-G\ (As3d)}}$ and $\frac{I_{Ga-Ga(Ga3d)}}{I_{Ga-A\ (As3d)}}$. Using SEM and AFM characterizations, the morphological parameters have been determined. Results are presented in the Table 3.

**Table 3.** XPS ratios of the nitridated samples and morphological parameters of the GaN clusters from SEM and AFM measurements.

| Sample | XPS ratios | | | GaN cluster parameter | | | |
|---|---|---|---|---|---|---|---|
| | $\frac{I_{Ga-Ga}}{I_{Ga-As}}$ | $\frac{I_{Ga3d}}{I_{As3d}}$ | $\frac{I_{Ga-N}}{I_{Ga-As}}$ | Dot density ($\times 10^{10} cm^{-2}$) | Diameter (nm) MEB | Diameter (nm) AFM | Height (nm) AFM |
| GaN150 | 0.1 | 1.4 | 0.48 | $11 \pm 2$ | $14 \pm 3$ | $16 \pm 3$ | $6 \pm 1$ |
| GaN300 | 0 | 1.7 | 0.56 | $2 \pm 1$ | $20 \pm 3$ | $22 \pm 3$ | $8 \pm 1$ |
| GaN300 + 500°C annealing | 0 | 1.8 | 0.6 | $3 \pm 1$ | $20 \pm 3$ | $26 \pm 3$ | $7 \pm 1$ |
| GaN100➔350 | 0 | 1.6 | 0.46 | $12 \pm 3$ | $16 \pm 3$ | $20 \pm 5$ | $6 \pm 1$ |
| GaN100➔350 + 500°C annealing | 0 | 1.63 | 0.5 | $13 \pm 2$ | $16 \pm 2$ | $20 \pm 4$ | $5 \pm 1$ |

The ratio $I_{Ga-Ga}/I_{Ga-As}$=0.1 indicates an incomplete nitridation of the Ga droplet, i.e. dissolution of the nitrogen atoms in the Ga droplet is not complete for GaN150. At this temperature an oxygen contamination as shown by the As-O bond in As3d and N-O bond in N1s is observed. Free arsenic $As^0$ is also observed. After nitridation at 150 °C, the density, diameter and height of the dots remain almost unchanged.

For GaN300, the Ga-Ga component is no longer observed in the Ga3d peak ($I_{Ga-Ga}/I_{Ga-As}$=0). Nitrogen diffusivity in gallium is sufficient at 300°C to completely consume the metallic gallium. Free $As^0$ and oxidized (As-O) arsenic in As3d and N-O bonds in N1s have a lower proportion than in the GaN150 sample due to the higher substrate temperature, favoring their desorption. These results are in agreement with those of H. Mehdi et al. in the reference [11] where desorption of $As^0$, As-O and N-O is shown, with increasing temperature. Nanostructure density decreases significantly to $2x10^{10}$ cm$^{-2}$, and the diameter and height of the nanostructures slightly increase to 22±3 nm and to 8±1 nm after nitridation, respectively. In addition, a thin amorphous GaN layer is formed on the substrate around the droplet. This layer is estimated in the HRTEM image (Fig.8(a)) to be around 2 nm thick. In this image, an interatomic distance of 0.33 nm in the GaAs(111)A substrate is observed which corresponds well to the GaAs lattice. The formation of well-defined atomic planes in the GaN nanostructure demonstrates the growth of crystalline GaN at 300°C. Indeed, the Fast-Fourier-transform (FFT) shows a zinc blende crystallinity with an interatomic distance of $d_{111}$=0.25 nm.

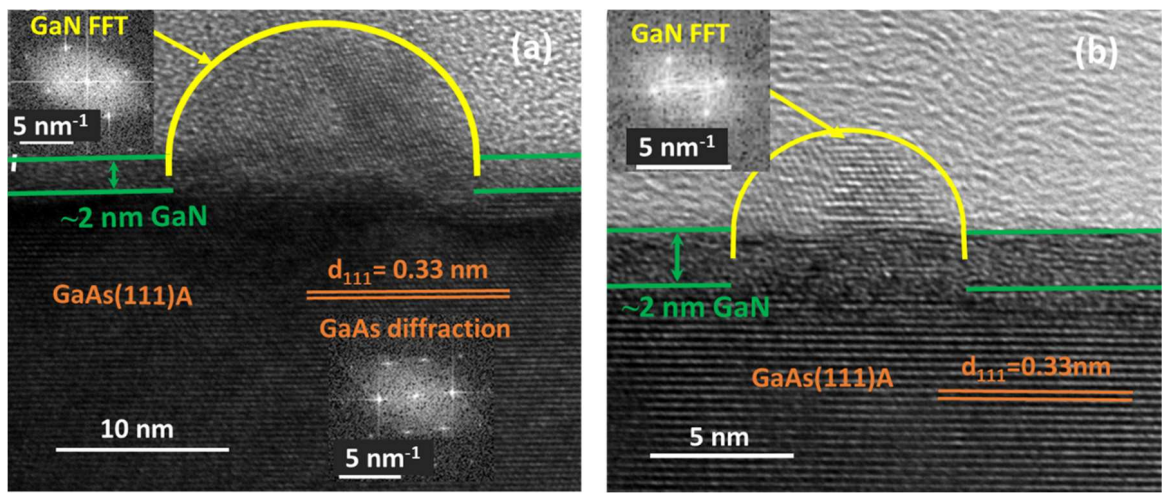

**Figure 8.** HRTEM image of (a) GaN300 and (b) GaN300 + annealing at 500 °C nanodots. Inserts show the diffraction patterns.

To reduce the oxygen contamination, an annealing at 500 °C has been performed on GaN300 + 500 °C annealing. After this process, the contributions of free arsenic As0 and As-O bonds in As3d and N-O bonds in N1s become almost zero. This is due to the annealing temperature, which desorbs them [11]. The $I_{Ga-Ga}/I_{Ga-As}$ XPS intensity ratio remains zero, as in the case of the GaN300 sample. Few changes are observed in the dot density ($3 \times 10^{10}$ cm$^{-2}$) and the diameter (26±2 nm) and the height (7±1 nm) after annealing. HRTEM analysis shows well-defined atomic planes in the zb-GaN nanostructures (Fig.8(b)) following the same orientation [111] as the substrate (epitaxial relation). The thin GaN layer on the GaAs(111)A substrate maintains a thickness of ~2 nm but appears to be better organized than on GaN300 due to the annealing at 500 °C.

From the AFM images, we mesure the center-to-center distance between the nearest droplets in all directions of space called $d_{min}$ and compute its mean value $\overline{d_{min}}$. The results obtained before and after the droplet nitridation are compared in the two rightmost panels of Fig.9. After deposition of metallic gallium, $\overline{d_{min}}$ is equal to 24 nm with a standard deviation equal to 6 nm (central panel, red dots in the online version). No preferential direction of diffusion of gallium atoms is apparent, as the distribution of the $d_{min}$ values is uniform as a function of the angle. After nitruration at 300 °C (GaN300 sample in the right hand panel, blue dots online), $\overline{d_{min}}$ increases to 155 nm and a larger relative variability of the minimum inter-droplet distances (±74 nm) is observed. When the sample is further annealed at 500 °C, consistently with the slight change in dot density mentioned above, $\overline{d_{min}}$ decreases slightly to 117±54 nm (right hand panel, green dots online) showing a reorganization of the GaN clusters.

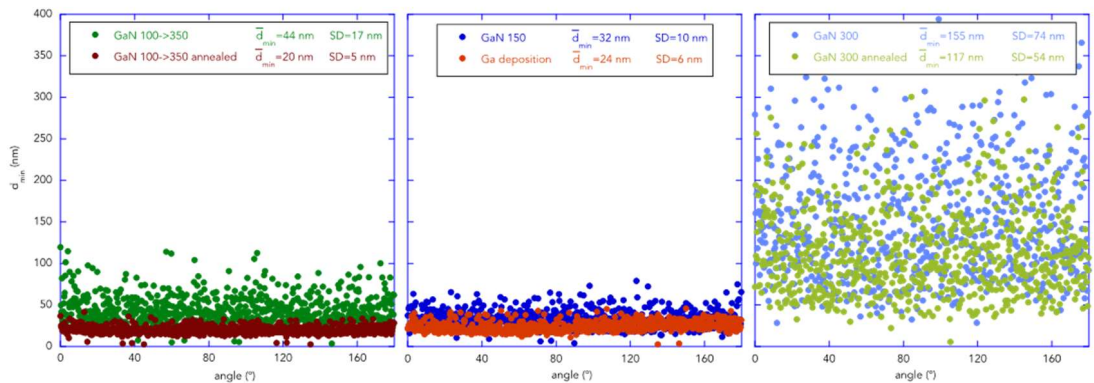

**Figure 9.** Distribution of $d_{min}$ after gallium deposition, after nitridation and after annealing.

In view to better understand the nitridation process, Ga droplet nitridation was analysed by KMC simulation. For nitridation we used the initial surface with the Ga drops obtained at $E_{diff}$ = 0.7 eV (Fig.5(b)). Nitridation was carried out at various temperatures with a nitrogen flux of $F_N$ = 0.1 ML/s (Fig.10).

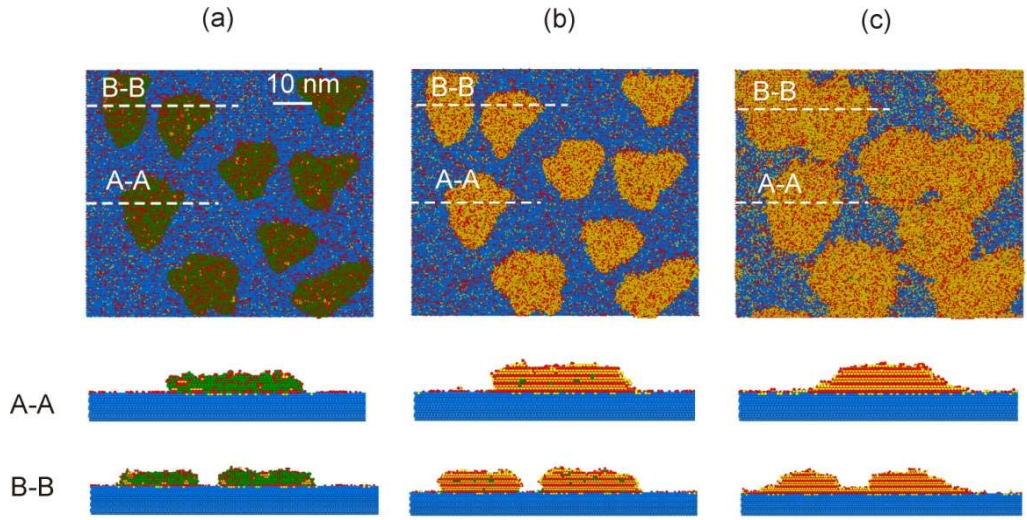

**Figure 60.** Top views and cross-section of GaAs(111)A with GaN clusters after nitridation for 1800 s at different temperatures: (a) T = 100 °C; (b) T = 200 °C; (c) T = 300 °C. GaAs substrate (blue colour), Ga (solid) yellow, Ga (liquid) green, nitrogen atoms are shown in red.

For nitridation conditions corresponding to Fig. 10(a), a slow GaN crystallization takes place on the drop-substrate interface and on the drop surface. Crystallization is not layer-by-layer due to low nitrogen diffusion in liquid droplet at low temperature. Presence of liquid gallium is in a good agreement with experimental observations of Ga-Ga bonds in Ga3d XPS peaks for sample GaN150.

For the conditions of Fig.10(b), almost complete droplet crystallization in the form of GaN dots occurs. At 200 °C, we observe the formation of Ga-N bonds, but this process is incomplete. There are defects such as Ga(liquid) inclusions and anti-site defects. This observation is consistent with XPS analysis of a sample nitridated at 200°C (not presented here) for which the recorded ratio $I_{Ga-Ga}/I_{Ga-As}$=0.08 indicates an incomplete nitridation of the Ga droplet.

For growth conditions of Fig.10(c) high quality GaN cluster is formed but the compact clusters spread slightly. Substrate cross-section (B-B) demonstrates GaN layer formation between closely spaced droplets. This layer is the result of N-atom interaction with Ga atoms diffusing from the droplet. A weak nitridation of the GaAs (111)A substrate can also be observed. This result is in good agreement with the results of sample GaN300. For nitridation at T=400 °C, GaN clusters overlap each other and thin GaN layer is formed between clusters (not shown here).

In Table 4, diameter, and height of model GaN clusters after Ga droplet nitridation are presented. One can see that cluster diameter slightly decreases in comparison with initial Ga droplet diameter. This is due to the lattice constant of cubic GaN that is lower than that of cubic GaAs. The mean height is higher than that of Ga droplets due to cluster volume increasing in comparison with droplet volume due to nitrogen dissolution into liquid gallium.

**Table 4.** Morphological characteristics of model GaN clusters after Ga droplet nitridation from KMC simulation.

| Temperature (°C) | 100 | 200 | 300 |
|------------------|-----|-----|-----|
| Diameter (nm) | 14.3 | 14.7 | 15.4 |
| Height (nm) | 2.2 | 2.5 | 3.2 |

In Fig.11, kinetics of Ga droplet nitridation at 100 °C and 300 °C is shown. At low temperature (Fig.11(a)), sub monolayer nitrogen coverage is formed on the substrate, but GaN cluster crystallization is not observed. However, after long nitridation time (t = 7200 s) core-shell structure is formed. The solid GaN shell formation needs time. After t=1800 s nitridation only GaN shell fragments are observed. At T=300 °C (Fig.11(b)), layer-by-layer GaN crystallization starts under the Ga droplets and at 200 s the drop is fully transformed into the GaN nanocluster.

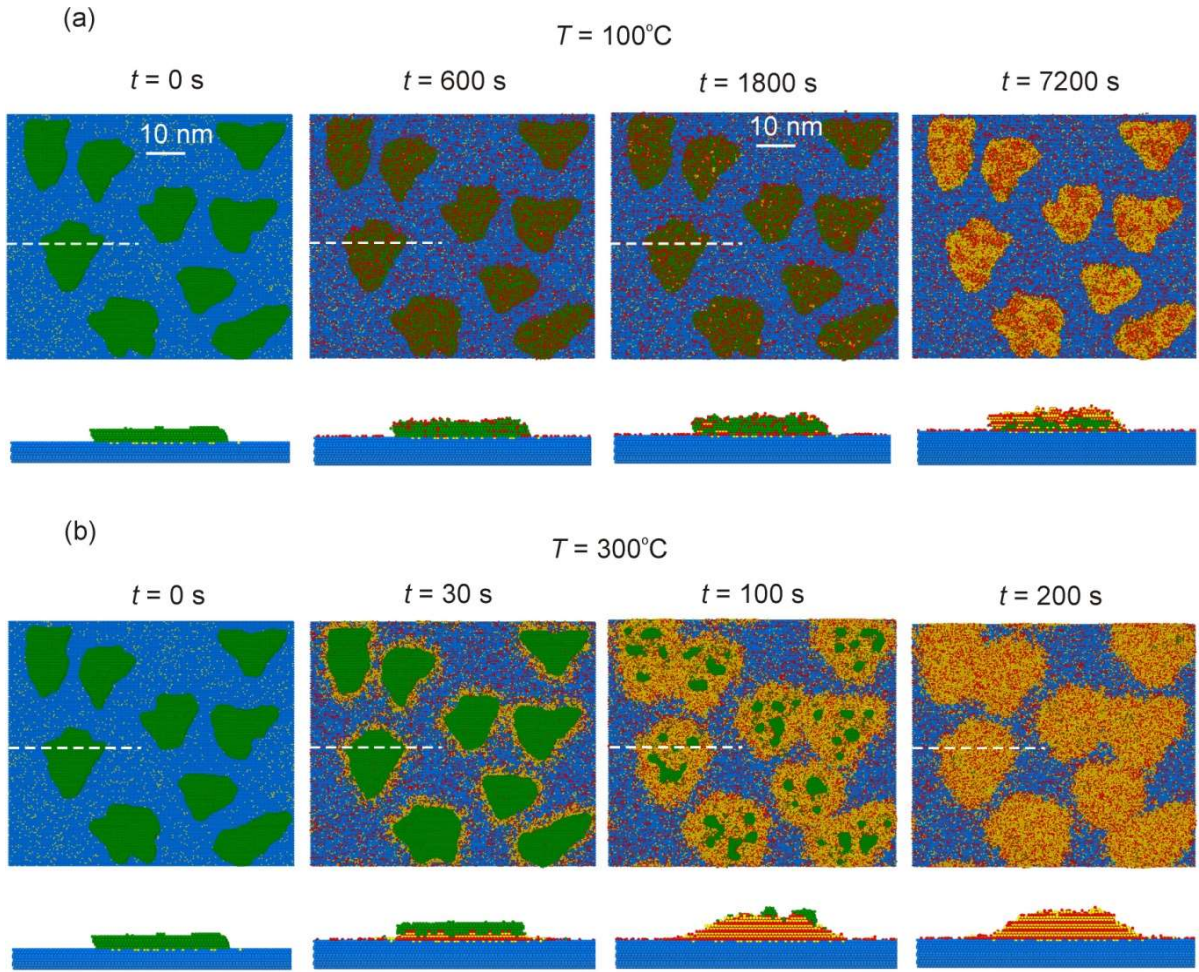

**Figure 7.** Kinetics of nitridation process at temperature (a) 100 °C and (b) 300 °C.

To summarize, during nitridation at low temperature (<200 °C) and high nitrogen flow, the density and morphology of the nanostructures are almost preserved. However, complete nitridation of the liquid metal drop is difficult because of the low rate of nitrogen diffusion in gallium under these conditions. Experiment and simulation demonstrate the presence of a liquid gallium phase inside the cluster, even at 200 °C. Simulation does not show the formation of core-shell structures at 200 °C, but only inclusions of liquid Ga phase inside GaN cluster. At these temperatures, surface diffusion is limited and therefore the cluster density remains constant at around $1\times10^{11}$ cm$^{-2}$. When the temperature is increased ($\geq$300 °C), a complete nitridation of the gallium droplets is predicted by KMC simulations and observed experimentally (disappearance of Ga-Ga bonds in the Ga3d peak in XPS). Crystalline GaN nanostructures are then grown. However, at this temperature, there is significant surface mass transport, which leads to a modification of the morphology and an undesirable reduction in the density ($2\times10^{10}$ cm$^{-2}$) of the GaN clusters at the end of the nitridation process.

As we have previously mentioned, annealing of the Ga droplets at 300 °C has no effect on the droplet density. The density change observed during nitridation must therefore be ascribed to the exposure to the nitrogen plasma. Computations show that the diffusion barrier of Ga adatoms is increased by the presence of excess N [40], so decreasing of cluster density during nitridation stage is unclear and requires additional research.

We conclude that in order to grow crystalline GaN nanostructures with a high density, two key phenomena must optimized: nitrogen diffusion in the Ga droplet; and surface mass transport between GaN clusters. Therefore, inspired by the work [41], we propose the following fabrication scheme: (i) we deposit liquid Ga droplets at 300 °C ; (ii) we lower the temperature to 100 °C to limit surface diffusion and maintain a high droplet density and switch on the N flux, then (iii) gradually increase the temperature up to 350 °C to complete nitridation of the gallium droplets. Finally, (iv) annealing the nitridated sample at 500 °C leads to a better homogeneity of the nanostructures.

This procedure yields the GaN100→350 sample. As for GaN300, Ga-Ga bonds are not observed in the Ga3d peak measured on GaN100→350, indicating a total droplet nitridation. The GaN cluster density ($1.2x10^{11}$ $cm^{-2}$), diameter (20±5 nm) and height (6±1 nm) are very close to the values for Ga droplets after Ga deposition (Table 1) because nitridation at low temperature hinders surface mass transport, as it can be seen for the GaN150 sample. Then, the gradual increase to 300-350 °C allows to speed up the nitridation process and to achieve completely nitridated GaN clusters, as demonstrated by the GaN300 sample. We can conclude that the nitridation with temperature ramp makes it possible to preserve the initial morphology of the gallium droplets as deposited.

After annealing (GaN100→350 + 500 °C annealing), the density of the nanostructures remains the same as after nitridation ($1.3x10^{11}$ $cm^{-2}$). The diameter and height of the GaN dots after annealing are almost unchanged (20±4 nm and 5±1 nm respectively). Note that annealing allows complete desorption of surface impurities ($As^0$, As-O and N-O). HRTEM (Fig.12) shows well-defined crystal planes in the GaN nanostructures. The diffraction pattern of a GaN nanodot exhibits a GaN zinc blende structure oriented in the same direction [111] as the substrate. The GaN layer formed in this sample (<1 nm-thick) is mainly due to the nitridation of the GaAs substrate. Indeed, hindering mass transport during nitridation strongly limits the formation of the ~~buffer~~ GaN layer on the GaAs substrate.

In Fig.9, $\overline{d_{min}}$ = 44±17 nm for GaN100→350 is much smaller than the minimum interdot distance found after a nitridation at 300 °C ($\overline{d_{min}}$ =155±74 nm). This shows that that the nanodot morphology is affected only by the first low temperature nitridation step and is not sensitive to the following steps, including annealing. After annealing, $\overline{d_{min}}$ = 20±5 nm, similar which is similar to the value before nitridation ($\overline{d_{min}}$ = 24±6 nm).

In conclusion, this process, exploiting a temperature ramp from 100°C to 350°C followed by a 500°C annealing, makes it possible to grow zb-GaN dots while maintaining a high cluster density of approximately 1x10$^{11}$ cm$^{-2}$.

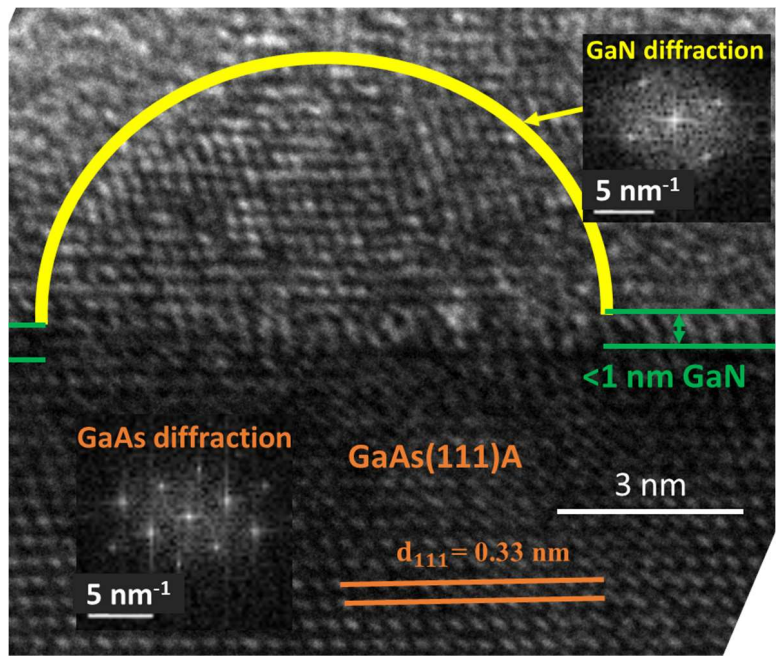

**Figure 8.** HRTEM image and diffraction of a GaN nanodot from the sample GaN100→350 + 500 °C annealing. Inserts show the diffraction patterns.

Results presented above should be further discussed as they differ from other III-V cluster growth on GaAs(111)A substrate. Previously, core-shell structures observed during droplet epitaxy of GaAs clusters were formed due to the diffusive inflow of As to the Ga drop from the substrate surface [13]. In this case formation of the GaAs shell starts from the contact line between the liquid droplet and the substrate (triple line). In our case, for low temperature (<200 °C), the diffusive inflow of nitrogen in both atomic and molecular form from the substrate is very small.

The diffusion of atomic nitrogen is small because of the high binding energy of nitrogen atoms with the GaAs(111)A substrate, and the concentration of $N_2$ molecules on the surface is small because they are very volatile. This is why another mechanism of shell formation is observed in KMC simulations (Fig.11(a)). At low temperature, isolated GaN nanocrystals are nucleated on the surface of a Ga droplet. Gradually these crystals coalesce and a continuous GaN shell forms. After complete core-shell formation, additional Ga droplet nitridation does not occur. At high temperature, surface diffusion and solubility in the Ga droplet of atomic nitrogen are enhanced. Shell formation is not observed. GaN crystal nucleation takes place from the GaAs/Ga droplet interface and complete nitridation of Ga droplets is obtained.

Based on the experimental results presented in Tables 1 and 3, the following conclusions can be drawn.

- At low temperature (GaN150), the surface density of Ga droplets equals that of GaN clusters, indicating negligible mass transport during this nitridation.

- Consistently, effective surface mass transport during nitridation at higher T (GaN300) lowers the density of GaN clusters considerably from initial density of Ga droplets, though it allows GaN clusters to be fully crystallized.

- Low temperature observations suggest that shell formation takes place on a shorter time scale than inter-cluster diffusion, due to hindering of surface mass transport between clusters ; this can be exploited by exposing Ga droplets formed at 300 °C to a nitrogen plasma at low substrate temperature and allowing a GaN shell to form on their outer surface.

- Ramping the temperature up to 350 °C allows N to diffuse through drop to the liquid-crystal interface and to achieve full crystallization of the GaN clusters, without any appreciable inter-cluster mass exchange.

- The result is an array of a fully crystallized GaN nanodots of density that is completely controlled by the initial liquid Ga droplets deposition step.

# 5 Conclusions

We have shown in this work how complementarity between KMC simulation and experimental characterizations (XPS, SEM, AFM, HRTEM) can be used to optimise the growth parameters of GaN nanodots on a GaAs(111)A substrate. The difficulty of nitridating gallium droplets at low temperatures (<200 °C) is due to the slow nitrogen diffusion in Ga liquid droplet. The complete nitridation of the Ga droplets was demonstrated by increasing the temperature at 300 °C. At this temperature, crystalline nanodots are observed by HRTEM. However, nitridating at 300 °C results in a decrease of the density of the GaN nanodots and the formation of a GaN layer of around 2 nm. In order to maintain the high density of the nanodots and to completely nitride the gallium droplets, we proposed nitridation during a ramping temperature, which consists of (i) starting nitridiation at a low temperature in order to decrease the surface mass transport during nitrogen plasma exposure and then (ii) gradually increasing the temperature in order to completely nitridate the gallium droplets. This nitridation process enable us to obtain a high density around $1 \times 10^{11}$ cm$^{-2}$ of zb-GaN nanocrystals which orientation is set by the substrate. This also reduces the GaN layer to < 1 nm-thick. Moreover, annealing at 500 °C induces a very good homogeneity of the GaN nanostructures and almost the complete desorption of surface impurities adsorbed during the nitridation.

**Corresponding Authors**

*Luc Bideux luc.bideux@uca.fr

*Guillaume Monier guillaume.monier@uca.fr

ACKNOWLEDGMENT

The authors gratefully acknowledge financial support from the Ministry of Higher Education and Research and Innovation supported by the French National Research Agency and of the Ministry of Education and Science of the Russian Federation. This work was sponsored by a public grant overseen by the French National Research Agency as part of the "Investissements d'Avenir" through the IMobS3 Laboratory of Excellence (ANR-10-LABX-0016) and the IDEX-ISITE initiative CAP 20-25 (ANR-16-IDEX-0001). The authors gratefully acknowledge

TOC GRAPHIC

**Figure 1.** The scheme of the model system and considered elementary events: 1 – particle adsorption; 2 – diffusion; 3 – nitrogen recombination $N + N \rightarrow N_2$; 4 – evaporation; 5 – Ga dissolution into liquid Ga; 6 – nitrogen dissolution and diffusion through the Ga droplet; 7 – GaN crystallization. GaAs substrate is marked by blue, Ga (solid) is marked by yellow and Ga (liquid) by green. Incoming nitrogen atoms are shown in red.

**Figure 2.** XPS peak decompositions after depositing 3 ML of Ga on GaAs (111)A.

**Figure 3.** AFM (a) and SEM (b) images taken on Ga/GaAs(111)A after 3ML Ga deposition at T=300 °C.

**Figure 4.** Droplet diameter (a) (from SEM) and height (b) (from AFM) distributions after 3 ML Ga depositing.

**Figure 5.** Monte Carlo simulation results: Model substrate top views and cross-sections along dashed line after 3 ML Ga deposition during 30 s on GaAs(111)A at T=300 °C as a function of Ga diffusion barrier energies $E_{diff}$: (a) 0.6 eV; (b) 0.7 eV; (c) 0.8 eV; (d) 0.9 eV. Substrate Ga(solid) and As(solid) atoms are marked in blue and cyan, respectively. Ga(liquid) is marked in green.

**Figure 6.** Energy as a function of the relative position of the diffusing gallium atom.

**Figure 7.** XPS peak decompositions after gallium droplet nitridation at 150 °C for 30 min (sample GaN150).

**Figure 8.** HRTEM image of (a) GaN300 and (b) GaN300 + annealing at 500 °C nanodots. Inserts show the diffraction patterns.

**Figure 9.** Distribution of $d_{min}$ after gallium deposition, after nitridation and after annealing.

**Figure 10.** Top views and cross-section of GaAs(111)A with GaN clusters after nitridation for 1800 s at different temperatures: (a) T = 100 °C; (b) T = 200 °C; (c) T = 300 °C. GaAs substrate (blue colour), Ga (solid) yellow, Ga (liquid) green, nitrogen atoms are shown in red.

**Figure 11.** Kinetics of nitridation process at temperature (a) 100 °C and (b) 300 °C.

**Figure 12.** HRTEM image and diffraction of a GaN nanodot from the sample GaN100➔350 + 500 °C annealing. Inserts show the diffraction patterns.

GRAPHICAL ABSTRACT

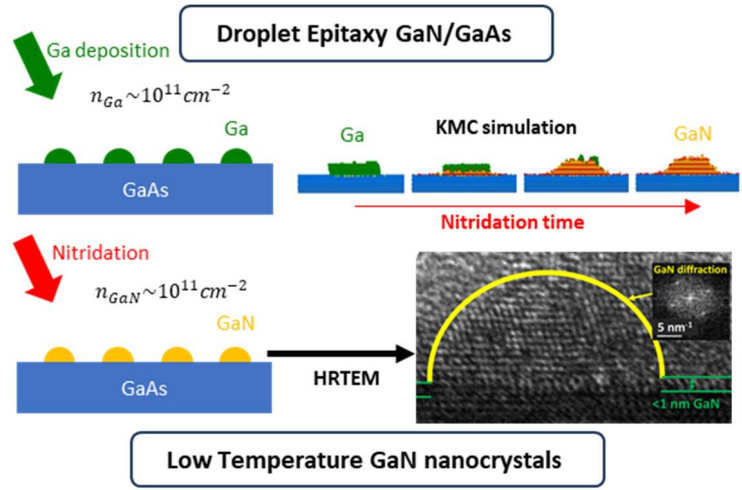

**Droplet Epitaxy GaN/GaAs**

Ga deposition

$n_{Ga} \sim 10^{11} cm^{-2}$

Ga

KMC simulation

GaN

Nitridation time

Nitridation

$n_{GaN} \sim 10^{11} cm^{-2}$

GaN

HRTEM

GaN diffraction

5 nm$^{-1}$

<1 nm GaN

**Low Temperature GaN nanocrystals**